\documentstyle[seceq,preprint]{ptptex}
\newcommand{\bra}[1]{\langle {#1} |}     %%
\newcommand{\ket}[1]{| {#1} \rangle}     %%
\newcommand{\lsim}{{\stackrel{<}{\sim}}}
\title{%        %You can use \\ for explicit line-break
On Parametric Resonance in Quantum Many-Body System}
\subtitle{Collective Motion and Quantum Fluctuation Around It in \\
Coupled Lipkin Model
}    %use this when you want a subtitle

\author{%       %Use \sc for the family name
Yasuhiko {\sc Tsue},$^{1}$ 
Jo\~ao da {\sc Provid\^encia},$^{2}$ 
Atsushi {\sc Kuriyama}$^{3}$ 
and \\
Masatoshi {\sc Yamamura}$^{3}$
}

\inst{%         %Affiliation, neglected when [addenda] or [errata]
$^{1}$Physics Division, Faculty of Science, Kochi University, Kochi 780-8520, 
Japan\\
$^{2}$Departamento de Fisica, Universidade de Coimbra, P-3000 Coimbra, 
Portugal\\
$^3$Faculty of Engineering, Kansai University, Suita 564-8680, Japan\\
}

\recdate{%      %Editorial Office will fill in this.
\today
}

\abst{%       %this abstract is neglected when [addenda] or [errata]
The dynamics governed by a requantized collective Hamiltonian 
in the coupled Lipkin model is investigated in the time-dependent 
variational approach with squeezed state. 
It is pointed out that there is a possibility of 
the parametric resonance mechanism which leads to amplifying 
the amplitude of quantum fluctuation around the collective mode 
in this model. 
}

\begin{document}

\maketitle

\section{Introduction}

The time-dependent Hartree-Fock (TDHF) theory and its approximated 
versions and/or the extensions have played a crucial role in the 
studies of nuclear collective motions. 
They are based on the time-dependent variational principle and exhibit 
two characteristic aspects. 
The aspect (i) is to give a possible description of time-evolution 
of the quantal state under investigation in the frame of a chosen 
form of the trial state of the variation. 
The aspect (ii) is to give a possible classical counterpart of the 
original quantal system under a suitable choice of the trial state. 
Then, it is expected that the original quantal system is 
reproduced in a disguised form under an appropriate requantization. 
The aspect (ii) has been deeply concerned in the studies of 
collective motions. 

Since a powerful idea was proposed by Marumori, Maskawa, Sakata and Kuriyama 
in 1980,\cite{MMSK80} 
various studies on the aspect (ii) have been performed until the present. 
For example, we can find the newest one in Ref.\citen{KNMM04}. 
At the early stage, the present authors (M. Y. and A. K.) also presented a 
possible form constructed in terms of canonical variables 
including the Grassmann variables, which was reviewed in Ref.\citen{YK87sup}. 
In order to make the significance of this paper understandable, first, 
we summarize the basic scheme of our form as follows:\\
(a) Paying attention to the Lie-algebraic structure, we set up a trial 
state containing parameters. 
These obey the canonicity condition\cite{YK87sup,KY84} 
and can be regarded as the canonical variables in classical mechanics. \\
(b) The trial state in (a) leads us to the classical counterpart 
of the original quantal systems through a certain procedure. 
It is formulated in the phase space of classical mechanics. \\
(c) In various cases, requantizations at this stage serve us quantal systems 
in disguised forms completely equivalent to the original ones, i.e., 
boson realization of Lie algebra. \\
(d) We presuppose that classical version for collective motions 
under investigation can be described on a collective submanifold 
in the phase space in (b). 
This is specified by canonical variables which enables us to describe the 
collective motion, i.e., 
collective variables. 
Under the above presupposition, equation of collective submanifold 
is derived. \\
(e) Together with the canonicity condition, the equation of collective 
submanifold enables us to express the variables in (a) as functions 
of the collective variables. 
Then, classical Hamiltonian is obtained in terms of the collective 
variables and by solving the Hamilton's equation of motion, 
we get the time-evolution of the original quantal system in the 
frame of the chosen form for the trial state. \\
(f) As the result of an appropriate requantization, we obtain a 
quantal system expressed in terms of the collective variables as the 
operators. Of course, it is of a disguised form for the original quantal 
system in the collective subspace. \\
(g) By diagonalizing the Hamiltonian in (e), the collective motion 
can be described. 
Naturally, we expect that the results are in good agreement with the 
exact one. \\
The above is our basic scheme reviewed in Ref.\citen{YK87sup}. 
The adiabatic TDHF approach can be formulated in this 
scheme.\cite{YK87sup,KY84} 
However, we must point out the following:\\
(h) The investigation of the aspect (i) based on the Hamiltonian in (f) 
has remained untouched. 
If, as the trial state, we adopt boson coherent state or its equivalent 
state for the Hamiltonian in (f), the time-dependent variation gives us 
the same results as those in (e) except the quantum effect coming from the 
ordering of operators. 
However, we have a possibility to adopt trial state different from the boson 
coherent state, for example, such as the squeezed state and we expect 
the aspect (ii) different from that in (e). 

With the aim of making a check on the validity of the above scheme 
((a)$\sim$(g)), the present authors (M. Y. and A. K.) with Iida 
investigated the adiabatic TDHF approximation on the 
coupled Lipkin model, a kind of the $su(2)\otimes su(2)$-algebraic model, 
which will be referred to as (A).\cite{KYI84} 
Pioneering result on this model can be seen in Ref.\citen{HI80}. 
By solving the equation of collective submanifold in the adiabatic TDHF 
approximation, we can draw the equi-potential curves in two dimensional space. 
In a certain region of the coupling strength, two bottoms appear 
and we can determine the collective path passing these two points by 
one parameter which plays a role of collective coordinate. 
After the procedure given in the scheme ((a)$\sim$(g)), 
we derive various results under rather good agreement with the exact one. 
Therefore, it may be interesting to investigate the aspect (i) 
in the sense of (h).

As was mentioned above, 
in (A), it was shown that the form of collective potential in the 
coupled Lipkin model has two 
minima in a certain region of coupling strength, that is, 
the collective potential is similar to the double well potential 
in the one dimensional problem in classical or quantum mechanics. 
If the collective variable oscillates around a minimum of the 
collective potential, it may be expected that the amplitude of this 
oscillation becomes gradually small because the oscillational energy 
dissipates to other degree of freedom such as quantum fluctuations or 
single particle motion. 
The similar situation has been realized theoretically in the late time of 
dynamical chiral phase transition in the context of the formation of 
disoriented chiral condensate.\cite{MM,HM} 
In this case, the chiral condensate oscillates around its vacuum value, 
where the chiral condensate corresponds to the collective mode in nuclear 
collective motion. 
Further, one of the present authors (Y.T.) has pointed out 
that, when the chiral condensate oscillates around the vacuum value 
in the late time of chiral phase transition in the context of the 
relativistic heavy ion collisions, there is 
a possibility that the amplitudes of 
quantum pion modes, which correspond to the 
quantum fluctuations around the mean field, are amplified 
by the mechanism of the parametric resonance and/or the forced oscillation 
induced by the oscillation of the chiral condensate in the 
O(4) linear sigma model.\cite{T02}
Also, the general investigation for the parametric resonance mechanism 
in the O(N) scalar field theory 
is given by using the $1/N$ expansion method.\cite{BS03} 

In this paper, as a part of the aspect (i) given in (h), 
we investigate the possibility of the parametric 
resonance mechanism to amplify the quantum fluctuation around the 
collective variable in the coupled Lipkin model developed in (A). 
It is important to investigate whether the amplification of fluctuation mode 
occurs or not. The reason is as follows : 
The amplification of the fluctuation mode may possibly 
lead to the damping of 
collective mode because the energy flow from collective motion 
to fluctuations should exist. 
In \S 2, the specification of collective submanifold in the coupled 
Lipkin model is recapitulated following (A). The derived 
collective Hamiltonian is requantized and we treat this system governed 
by the collective Hamiltonian as 
a quantum mechanical system. In \S 3, 
the time-dependent variational approach with squeezed state is applied 
to the dynamical problem of collective motion including quantum fluctuation on 
the collective submanifold. 
In \S 4, it is shown that the time-dependent part of collective mode 
induces the parametric resonance for the quantum fluctuation mode, in which 
it is demonstrated that the Mathieu equation with additional term is derived. 
In \S 5, discussion and concluding remarks are presented.

\section{Recapitulation of specification of collective submanifold in the 
coupled Lipkin model}

In this section, we present a brief review about a specification of collective 
submanifold of the coupled Lipkin model 
by the adiabatic time-dependent Hartree-Fock method given in (A). 

The Hamiltonian of the coupled Lipkin model is given by 
\begin{subequations}\label{2-1}
\begin{eqnarray}
& &{\hat H}=\sum_{\sigma=1}^2{\hat H}_{\sigma}
-V_3({\hat S}_{1+}{\hat S}_{2-}+{\hat S}_{2+}{\hat S}_{1-}) \ , 
\label{2-1a}\\
& &{\hat H}_{\sigma}=2\epsilon_{\sigma}{\hat S}_{\sigma 0}
-\frac{1}{2}V_{\sigma}({\hat S}_{\sigma +}^2+{\hat S}_{\sigma -}^2) \ , 
\label{2-1b}
\end{eqnarray}
\end{subequations}
where the quasi-spin operators are defined in terms of the particle and 
hole creation and annihilation operators, $({\hat a}_{\sigma j m}^*, 
{\hat b}_{\sigma j m}^*)$ and $({\hat a}_{\sigma j m}, 
{\hat b}_{\sigma j m})$ :
\begin{eqnarray}\label{2-2}
& &{\hat S}_{\sigma +}=\sum_{m=-j}^j{\hat a}_{\sigma j m}^*(-)^{j-m}
{\hat b}_{\sigma j {\tilde m}}^* \ , 
\qquad
{\hat S}_{\sigma -}={\hat S}_{\sigma +}^{\dagger}\ , \nonumber\\
& &{\hat S}_{\sigma 0}=\frac{1}{2}\sum_{m=-j}^j
({\hat a}_{\sigma j m}^*{\hat a}_{\sigma j m}
+{\hat b}_{\sigma j m}^*{\hat b}_{\sigma j m})-\Omega\ , \nonumber\\
& &\Omega=j+\frac{1}{2}\ .
\end{eqnarray}
In this paper, we consider only the case of zero seniority number. 
Thus, the classical image of this system described by (\ref{2-1}) 
can be obtained by the TDHF method where the Slater determinantal state 
is used. As a result, a classical correspondence of 
the above quasi-spin operators is obtained 
in the following forms in terms of a certain set of canonical variables 
$(q^{\sigma},p_{\sigma})$ $(\sigma=1,2)$ :\cite{KY83} 
\begin{eqnarray}\label{2-3}
& &S_{\sigma x}=\frac{1}{2}(S_{\sigma +}+S_{\sigma -})
=\sqrt{\Omega^2-p_{\sigma}^2}\ \sin q^{\sigma} \ , \nonumber\\
& &S_{\sigma y}=\frac{1}{2i}(S_{\sigma +}-S_{\sigma -})
=-p_{\sigma} \ , \nonumber\\
& &S_{\sigma z}=S_{\sigma 0}=-\sqrt{\Omega^2-p_{\sigma}^2}\ \cos q^{\sigma}
\ . 
\end{eqnarray}
Thus, the Hamiltonian is reduced to the classical Hamilton function as 
\begin{eqnarray}\label{2-4}
& &H=\sum_{\sigma=1}^2 H_{\sigma}-V_3(S_{1+}S_{2-}+S_{2+}S_{1-}) \ , 
\nonumber\\
& &H_{\sigma}=-\biggl[2\epsilon_{\sigma}\sqrt{\Omega^2-p_{\sigma}^2}\ 
\cos q^\sigma 
%\nonumber\\
%& &\qquad\qquad\qquad
+\frac{V_{\sigma}}{\Omega}\left(\Omega-\frac{1}{2}\right)
%(1+\sin^2 q^{\sigma})
\{(\Omega^2-p_{\sigma}^2)\sin^2 q^{\sigma}-p_{\sigma}^2\}
\biggl] \ . \qquad
\end{eqnarray}

For the collective variables $(Q,P)$, 
we can derive the equation of collective submanifold 
and the canonicity condition following 
Ref.\citen{KY84}: 
\begin{eqnarray}
& &\partial_{q^{\sigma}}H
=\lambda\partial_P p_{\sigma}-\mu\partial_Q p_{\sigma} \ , \nonumber\\
& &\partial_{p_{\sigma}}H
=-\lambda\partial_P q^{\sigma}+\mu\partial_Q q^{\sigma} \ ,
\label{2-5}\\
& &\sum_{\sigma=1}^2 p_{\sigma}\partial_Q q^{\sigma}=P \ , \qquad
\sum_{\sigma=1}^2p_{\sigma}\partial_{P}q^{\sigma}=0 \ . 
\label{2-6}
\end{eqnarray} 
Here, $\partial_z=\partial/\partial z$. 
Thus, the basic equations consist of Eqs.(\ref{2-5}) and (\ref{2-6}). 

To proceed the calculation concretely, we introduce the adiabatic 
approximation which corresponds to the lowest approximation in terms of 
a power series of $P$. 
Under this approximation, 
$q^{\sigma}$ and $p_{\sigma}$ can be expressed as 
\begin{equation}\label{2-7}
q^{\sigma}=q^{\sigma}(Q) \ , \qquad p_{\sigma}=p_{\sigma}(Q)P \ .
\end{equation}
In this approximation, the Hamiltonian (\ref{2-4}) takes the following 
form : 
\begin{subequations}\label{2-8}
\begin{eqnarray}
& &H=\frac{1}{2}\sum_{\sigma,\sigma'=1}^2M^{\sigma\sigma'}p_{\sigma}p_{\sigma'}
+V(q^{\sigma}) \ , 
\label{2-8a}\\
& &M^{\sigma\sigma}=\frac{1}{\Omega}\left[2\epsilon_{\sigma}\cos q^{\sigma}
+2V_{\sigma}\left(\Omega-\frac{1}{2}\right)(1+\sin^2 q^{\sigma})\right]
+2V_3\sin q^1 \sin q^2 \ , \nonumber\\
& &M^{12}=M^{21}=-2V_3 \ , 
\label{2-8b}\\
& &V(q^{\sigma})=-\sum_{\sigma=1}^2 \Omega\left[
2\epsilon_{\sigma} \cos q^{\sigma}+V_{\sigma}\left(\Omega-\frac{1}{2}\right)
\sin^2 q^{\sigma}\right]-2V_3{\Omega}^2\sin q^1 \sin q^2 \ . \nonumber\\
& &
\end{eqnarray}
\end{subequations}
%where we have defined $\Omega'=\Omega+1/2$. 
Then, the basic equations (\ref{2-5}) and (\ref{2-6}) are reduced to 
the following equations : 
\begin{subequations}\label{2-9}
\begin{eqnarray}
& &C(Q)Q\cdot p_{\sigma}(Q)=\partial_{q^{\sigma}}V \ , \nonumber\\
& &\frac{dq^{\sigma}(Q)}{dQ}
=\sum_{\sigma'=1}^2 M^{\sigma\sigma'}p_{\sigma'}(Q) \ , 
\label{2-9a}\\
& &\sum_{\sigma=1}^2 p_{\sigma}(Q)\frac{dq^{\sigma}(Q)}{dQ}=1 \ . 
\label{2-9b}
\end{eqnarray}
\end{subequations}
The second of Eq.(\ref{2-6}) is automatically satisfied for (\ref{2-7}).
The first of Eq.(\ref{2-6}) is nothing but Eq.(\ref{2-9b}). 
The equations of collective submanifold (\ref{2-5}) are reduced to 
Eq.(\ref{2-9a}) with the explicit expressions of the Lagrange 
multipliers.\cite{KY84} Here, we do not need to know the explicit form 
of $C(Q)$. 

It should be here noted that the basic equations (\ref{2-5}) and (\ref{2-6}) 
do not give a unique collective coordinate system because these equations 
are still invariant with respect to the point canonical transformation 
$Q'=Q'(Q)$ and $P'=(dQ/dQ')\cdot P$. In (A), a possible method to avoid 
this ambiguity has been proposed. 
According to (A), the mass of collective motion has to be taken as unit 
in order to fix the collective coordinate system. 
%Then, from Eq.(\ref{2-9}) with the requirement of the mass of 
%collective motion being one, the following relation should be satisfied : 
In our case, (\ref{2-9a}) and (\ref{2-9b}), the mass of collective motion, 
$M
=(\sum_{\sigma,\sigma'=1}^{2}
p_\sigma(Q) M^{\sigma\sigma'}p_{\sigma'}(Q))^{-1}$, 
is automatically one from the second equation of (\ref{2-9a})and (\ref{2-9b}). 
Further, by multiplying $dq^{\sigma}(Q)/dQ$ on both sides of 
the first equation of (\ref{2-9a}) and summing up $\sigma=1,2$, 
it is shown that the 
following relation should be satisfied: 
\begin{equation}\label{2-10}
C(Q)Q=\partial_Q V \ .
\end{equation}
Thus, the collective Hamilton function is given by 
\begin{eqnarray}\label{2-11}
& &H_C=\frac{1}{2}P^2+V_C(Q) \ , \nonumber\\
& &V_C(Q)=\int^Q C(Q')Q'dQ' \ . 
\end{eqnarray}
Equations (\ref{2-9}) can be solved in the following power series expansion 
technique : 
\begin{eqnarray}\label{2-12}
q^{\sigma}(Q)=\sum_{n=0}q_n^{\sigma} Q^{2n+1} \ , \quad
p_{\sigma}(Q)=\sum_{n=0}p_{\sigma}^n Q^{2n} \ , \quad
C(Q)=\sum_{n=0}c_n Q^{2n} \ . \quad
\end{eqnarray}
It is interesting to the collective potential energy $V_C(Q)$ because 
we investigate the dynamics of collective motion and 
of the quantum fluctuation around the 
collective variable in this paper. 
In the above expansion, the collective potential $V_C(Q)$ 
can be represented as 
\begin{equation}\label{2-13}
V_C(Q)=\frac{1}{2}c_0 Q^2+\frac{1}{4}c_1 Q^4+\frac{1}{6}c_2 Q^6+\cdots 
+{\rm const.}
\end{equation}

Here, we numerically 
estimate the coefficients of power expansion in (\ref{2-12}). 
In the numerical evaluation, we adopt the following set of values for 
the model parameters as 
\begin{eqnarray}\label{2-14}
& &\epsilon_1=1.5 \ , \qquad \epsilon_2=2.0 \ , \nonumber\\
& &V_1=0.05\chi\ , \qquad V_2=0.1\chi \ , \qquad V_3=0.075\chi\ , 
\nonumber\\
& &\Omega=5 \ , 
\end{eqnarray}
which are those used in Ref.\citen{HI80} and in (A). 
Here, $\chi$ remains as a parameter which controls the force strength. 
The numerical values for $c_0$, $c_1$ and $c_2$ in the collective potential 
are listed up in Table I for the various $\chi$. 
It is found that, in the neighborhood of the phase transition point, that is 
around $c_0=0$, the collective potential can be approximated 
up to the order of $Q^4$ safely because the relation between the coefficients
of ${Q}^4$ and ${Q}^6$ satisfies $|c_2/6|/|c_1/4| \lsim 0.1$.

\begin{table}[t]
\caption{The values of coefficients, $c_0$, $c_1$ and $c_2$, 
of the collective potential $V_C=(1/2)c_0 Q^2+(1/4)c_1 Q^4 + (1/6)c_2 Q^6$ 
are listed. 
}
\label{table:1}
\begin{center}
\begin{tabular}{c|ccc|c} \hline \hline
$\chi$ & $c_0$ & $c_1$ & $ c_2$ & $|(1/6)c_2|/|(1/4)c_1|$ \\ \hline 
1.5 & 4.69 & 0.328 & $-$0.161 & 0.3278 \\
2.0 & 2.77 & 0.894 & $-$0.175 & 0.130 \\
2.5 & 0.942 & 1.24 & $-$0.106 & 0.0568 \\
3.0 & $-$0.747 & 1.40 & $-$0.002 & 9.5$\times$10$^{-4}$ \\
3.5 & $-$2.26 & 1.38 & 0.125 & 0.0602 \\
4.0 & $-$3.56 & 1.20 & 0.265 & 0.147 \\
4.5 & $-$4.62 & 0.881 & 0.403 & 0.305 \\
5.0 & $-$5.40 & 0.441 & 0.522 & 0.789 \\
\hline
\end{tabular}
\end{center}
%\end{wraptable}
\end{table}

We quantize 
the collective Hamiltonian in (\ref{2-11}) 
by the 
canonical quantization : $[\ {\hat Q}\ , \ {\hat P}\ ]=i\hbar$. 
In the next section, we will investigate the dynamics of the collective 
motion and the quantum fluctuation around it based on the above 
derived collective Hamiltonian up to the order of ${\hat Q}^4$.

%%%%%%%%%%%%%%%%%%%%%%%%%%%%% (Fig.1) %%%%%%%%%%%%%%%%%%%%%%%%%%%%%%%
%\begin{figure}[t]
%  \epsfxsize=13cm  % \epsfysize=   cm  
%  \centerline{\epsfbox{potential.eps}}
%  \caption{ }
%   \label{fig:1}
%\end{figure}
%%%%%%%%%%%%%%%%%%%%%%%%%%%%%%%%%%%%%%%%%%%%%%%%%%%%%%%%%%%%%%%%%%%%

\section{Time-dependent variational approach with squeezed state 
to the coupled Lipkin model on the collective submanifold}

We apply the time-dependent variational approach with a squeezed state 
to the coupled Lipkin model on the collective submanifold. 
Our task is reduced to solving the dynamical problem in one-dimensional 
quantum mechanical system. 
The time-dependent variational method with the squeezed state 
gives a useful approximation including the quantum effects in 
various quantal systems.\cite{TFKY91,TF91,KPTYsuppl}

\subsection{Squeezed state approach}
The squeezed state is defined as 
\begin{equation}\label{3-1}
\ket{\psi(\alpha,\beta)}=(1-\beta^*\beta)^{\frac{1}{4}}
\exp\left(\frac{\beta}{2}{\hat b}^{*2}\right)\cdot
\exp\left(-\frac{1}{2}\alpha^*\alpha\right)\exp(\alpha{\hat a}^*)
\ket{0} \ .
\end{equation}
Here, ${\hat a}^*$ is a boson creation operator and 
$\ket{0}$ is a vacuum state for the boson annihilation 
operator ${\hat a}$ : ${\hat a}\ket{0}=0$. 
The operators ${\hat b}^*$ and ${\hat b}$ 
are defined as 
\begin{equation}\label{3-2}
{\hat b}^*={\hat a}^*-\alpha^* \ , \qquad
{\hat b}={\hat a}-\alpha \ .
\end{equation}
The operator ${\hat b}$ is identical with the annihilation 
operator for the usual coherent state:
${\hat b}
\exp\left(-\frac{1}{2}\alpha^*\alpha\right)\exp(\alpha{\hat a}^*)
\ket{0}=0$. 
Introducing the operators which correspond to the coordinate and 
momentum operators, we have another expression of the squeezed state 
(\ref{3-1}) :
\begin{eqnarray}
\ket{\psi(\alpha,\beta)}
&=&e^{i\varphi}
(2G)^{-1/4}\exp\left(\frac{i}{\hbar}(P{\hat Q}-Q{\hat P})\right)
\exp\left\{\frac{1}{2\hbar}\left(
1-\frac{1}{2G}+i2\Pi\right){\hat Q}^2\right\}\ket{0} \ , 
\nonumber\\
& & \label{3-3}\\
& &{\hat Q}=\sqrt{\frac{\hbar}{2}}({\hat a}^*+{\hat a}) \ , \qquad
\qquad
{\hat P}=i\sqrt{\frac{\hbar}{2}}({\hat a}^*-{\hat a}) \ , 
\label{3-4}\\
& &{Q}=\sqrt{\frac{\hbar}{2}}({\alpha}^*+{\alpha}) \ , \qquad
\qquad
{P}=i\sqrt{\frac{\hbar}{2}}({\alpha}^*-{\alpha}) \ , 
\nonumber\\
& &G=\left| \sqrt{\frac{1}{2}+|y|^2}+y \right|^2 \ , 
\qquad
\Pi=\frac{i}{2}(y^*-y)\sqrt{\frac{1}{2}+|y|^2}\ G^{-1} 
\ , \label{3-5}\\
& &e^{-i2\varphi}
=\frac{1}{\sqrt{G}}\left(\sqrt{\frac{1}{2}+|y|^2}+y\right)
\ , \nonumber
\end{eqnarray}
where $y$ is related to $\beta$ as 
\begin{equation}\label{3-6}
y=\beta/\sqrt{2(1-|\beta|^2)}\ .
\end{equation}
The reason why we have introduced 
the new variables $y$ and $y^*$ is that these variables correspond to the 
boson-type canonical variables.
The expectation values for the coordinate and the momentum operators 
are derived easily as 
\begin{eqnarray}
& &
\bra{\psi(\alpha,\beta)}{\hat Q}\ket{\psi(\alpha,\beta)}
=Q \ , \nonumber\\
& &\bra{\psi(\alpha,\beta)}{\hat P}\ket{\psi(\alpha,\beta)}
=P \ , \nonumber\\
& &\bra{\psi(\alpha,\beta)}{\hat Q}^2\ket{\psi(\alpha,\beta)}
=Q^2+\hbar G \ , \nonumber\\
& &\bra{\psi(\alpha,\beta)}{\hat P}^2\ket{\psi(\alpha,\beta)}
=P^2+\hbar \left(\frac{1}{4G}+4G\Pi^2\right) \ , \quad \ \ \ \ 
\label{3-7}\\
& &\bra{\psi(\alpha,\beta)}V({\hat Q})\ket{\psi(\alpha,\beta)}
=
\exp\left\{\frac{1}{2}\hbar G\left(\frac{\partial}{\partial Q}\right)^2
\right\} V(Q) \ , 
\label{3-8}\\
& &\bra{\psi(\alpha,\beta)}\partial_z\ket{\psi(\alpha,\beta)}
=\frac{i}{2\hbar}(Q\partial_z P-P\partial_z Q)+iG\partial_z \Pi 
+i\partial_z \varphi \ , 
\label{3-9}
\end{eqnarray}
where $\partial_z=\partial/\partial z$.

The squares of the standard deviations for ${\hat Q}$ and ${\hat P}$ 
are then expressed as 
$\bra{\psi(\alpha,\beta)}({\hat Q}\!-\!Q)^2\ket{\psi(\alpha,\beta)}
=\hbar G$ and 
$\bra{\psi(\alpha,\beta)}({\hat P}\!-\!P)^2\ket{\psi(\alpha,\beta)}
=\hbar (1/(4G)+4G\Pi^2)$. 
Thus, the uncertainty
relation is expressed in terms of $G$ and $\Pi$ 
as 
$\bra{\psi(\alpha,\beta)}({\hat Q}-Q)^2\ket{\psi(\alpha,\beta)}
\bra{\psi(\alpha,\beta)}({\hat P}-P)^2\ket{\psi(\alpha,\beta)}
=\hbar^2(1/4+4G^2\Pi^2)$. 
It can be seen from this uncertainty relation that, 
if one direction of the uncertainty is relaxed, 
the other can be squeezed. 
Also, one can see from (\ref{3-8}) that the quantum effects beyond 
the order of $\hbar$ are included in this formalism. 
This novel feature is originated from the degree of freedom of 
the squeezing, namely $\beta$ and $\beta^*$ in (\ref{3-1}) or 
$G$ and $\Pi$ in (\ref{3-3}).

The time-evolution of this quantum state is governed by 
the time-dependent variational principle :
\begin{equation}\label{3-10}
\delta\int_{t_0}^{t_1}dt \bra{\psi(\alpha,\beta)}
i\hbar\partial_t - {\hat H} \ket{\psi(\alpha,\beta)} = 0 \ .
\end{equation}
In order to formulate the time-dependent variational approach 
in the canonical form, we impose the canonicity conditions as 
\begin{eqnarray}\label{3-11}
& &\bra{\psi(\alpha,\beta)} i\hbar \partial_{Q_i} \ket{\psi(\alpha,\beta)}
=P_i+\partial_{Q_i} s(Q_i,P_i) \ , \nonumber\\
& &\bra{\psi(\alpha,\beta)} i\hbar \partial_{P_i} \ket{\psi(\alpha,\beta)}
=\partial_{P_i} s(Q_i,P_i) \ , \quad (i=1,2) \ .
\end{eqnarray}
A set of possible solutions are obtained as 
\begin{eqnarray}\label{3-12}
& &Q_1=Q\ , \qquad P_1=P \ , \nonumber\\
& &Q_2=\hbar G \ , \qquad P_2=\Pi \ , \qquad s=-PQ/2+\hbar G\Pi +\hbar \varphi
\ . 
\end{eqnarray}
Thus, the equations of motion are formulated as canonical equations of 
motion with the same form in the classical mechanics : 
\begin{eqnarray}
& &{\dot Q}=\frac{\partial \langle {\hat H} \rangle}{\partial P} =P \ , 
\qquad\qquad 
{\dot P}=-\frac{\partial \langle {\hat H} \rangle}{\partial Q} \ , 
\label{3-13}\\
& &\hbar{\dot G}=\frac{\partial \langle {\hat H} \rangle}{\partial \Pi} 
= \hbar\cdot 4G\Pi \ , 
\qquad 
\hbar{\dot \Pi}=-\frac{\partial \langle {\hat H} \rangle}{\partial G} \ , 
\label{3-14}
\end{eqnarray}
where the dot denotes the time-derivative and $\langle {\hat H} \rangle$ 
is the expectation value of the Hamiltonian with respect to the state 
(\ref{3-1}) or (\ref{3-3}).

\subsection{Squeezed state approach to the 
coupled Lipkin model with the double well potential 
on the collective submanifold}

Let us consider the case in which the collective Hamiltonian 
of the coupled Lipkin model derived in \S 2 has a form of the 
double well potential, for example, in the region around $\chi \sim 3.5$. 
In this case, the Hamiltonian is simply written as 
\begin{equation}\label{4-1}
{\hat H}=\frac{1}{2}{\hat P}^2+\frac{1}{2}c_0 {\hat Q}^2 + 
\frac{1}{4}c_1 {\hat Q}^4 \ 
\end{equation}
with $c_0 < 0$ and $c_1 >0$, 
where the constant term in the collective potential is omitted because 
of no influence of the dynamics discussed later. 
The expectation value of this Hamiltonian with respect to the squeezed state 
is calculated as 
\begin{eqnarray}\label{4-2}
\langle {\hat H} \rangle &=&
\frac{1}{2}\left[P^2+\hbar\left(\frac{1}{4G}+4G\Pi^2\right)\right]
+\frac{1}{2}c_0(Q^2+\hbar G) \nonumber\\
& &+\frac{1}{4}c_1(Q^4+6\hbar Q^2 G+3\hbar^2 G^2) \ .
\end{eqnarray}
Then, the equations of motion in (\ref{3-13}) and (\ref{3-14}) are 
summarized as 
\begin{eqnarray}\label{4-3}
& &{\dot Q}=P \ , \nonumber\\
& &{\dot P}=-\left(c_0+c_1 Q^2+3\hbar c_1 G\right)Q \ , \nonumber\\
& &\hbar {\dot G}=\hbar \cdot 4G\Pi \ , \nonumber\\
& &\hbar{\dot \Pi}=-\hbar\left(-\frac{1}{8G^2}+2\Pi^2+\frac{1}{2}c_0
+\frac{3}{2}c_1 Q^2+\frac{3}{2}c_1\hbar G\right) \ .
\end{eqnarray}
Eliminating $p$ and $\Pi$, the above equations of motion are rewritten as 
\begin{eqnarray}
& &{\ddot Q}+\left(c_0+c_1 Q^2+3\hbar c_1 G\right)Q=0 \ , 
\label{4-4}\\
& &\frac{{\ddot G}}{4G}-\frac{{\dot G}^2}{8G^2}-\frac{1}{8G^2}
+\frac{1}{2}c_0+\frac{3}{2}c_1 Q^2+\hbar\cdot\frac{3}{2}G
=0 \ .
\label{4-5}
\end{eqnarray}
Further, since $G$ is positive definite, we introduce new variables 
$\eta$ instead of $G$ as 
\begin{equation}\label{4-6}
G=\eta^2 \ . 
\end{equation}
Then, the equations of motion (\ref{4-4}) and (\ref{4-5}) are further 
recast into 
\begin{eqnarray}
& &{\ddot Q}+\left(c_0+c_1 Q^2+\hbar\cdot 3c_1 \eta^2\right)Q=0 \ , 
\label{4-7}\\
& &{\ddot \eta}+\left(
c_0+3c_1 Q^2+\hbar\cdot 3c_1 \eta^2\right)\eta
-\frac{1}{4\eta^3}=0 \ .
\label{4-8}
\end{eqnarray}
Here, $Q$ represents the collective coordinate which presents a classical 
image of the collective motion. 
Also, $\eta$ represents quantum fluctuations around the collective 
variables found in Eq.(\ref{3-7}) with (\ref{4-6}).

\section{Generalized Mathieu's equation and 
amplification of quantal fluctuation mode by 
the parametric resonance mechanism}

First, we wish to derive the static solutions of the equations 
of motion (\ref{4-7}) and (\ref{4-8}). 
We denote the static solutions of $Q$ and $\eta$ as $Q_0$ and $\eta_0$, 
respectively. 
Then, $Q_0$ for $Q_0\neq 0$ and $\eta_0$ satisfy 
\begin{eqnarray}\label{5-1}
& &c_0+c_1{Q_0^2}+\hbar\cdot 3c_1{\eta_0^2}=0 \ , 
\nonumber\\
& &\left(c_0+3c_1{Q_0^2}+\hbar\cdot 3c_1{\eta_0^2}\right)
\eta_0^4=\frac{1}{4} \ . 
\end{eqnarray}
If $\hbar\rightarrow 0$ at this stage, the static solutions are easily 
derived as 
\begin{eqnarray}\label{5-2}
& &Q_0^2=-\frac{c_0}{c_1}\ (>0) \ , \nonumber\\
& &\eta_0^4=-\frac{1}{8c_0}\ (>0) \ . 
\end{eqnarray}

Next, let us investigate the time-dependent solutions $Q$ and $\eta$. 
We are restricted ourselves to seek the time-dependent solutions 
around the static configuration. Thus, 
the variables $Q$ and $\eta$ can be expanded as 
\begin{equation}\label{5-3}
Q=Q_0+\delta Q \ , \qquad \eta=\eta_0+\delta \eta \ . 
\end{equation}
Here, $\delta Q$ and $\delta\eta$ have time-dependence and 
we assume that these are small deviation. 
Substituting (\ref{5-3}) into (\ref{4-7}) and (\ref{4-8}) and 
using the relation (\ref{5-1}), we can obtain the following equations 
up to the order of $\delta Q$ and $\delta\eta$ : 
\begin{eqnarray}
& &\delta{\ddot Q}+\left(2c_1 Q_0^2
+\hbar\cdot 6c_1 \eta_0\delta\eta\right)\delta Q = 
-\hbar\cdot 6c_1 Q_0\eta_0\delta\eta \ , 
\label{5-4}\\
& &\delta{\ddot \eta}+\left(\frac{1}{\eta_0^4}+6c_1 Q_0\delta Q
+\hbar\cdot 6c_1 \eta_0^2\right)\delta\eta
=-6c_1 Q_0\eta_0\delta Q \ , 
\label{5-5}
\end{eqnarray}
where we have neglected the terms with $\delta Q^2$ and $\delta\eta^2$ 
and their higher order terms. 
If the semi-classical limit is adopted at this stage, namely 
$\hbar\rightarrow 0$, then we obtain 
\begin{eqnarray}
& &\delta{\ddot Q}-2c_0\delta Q \approx 0 \ , 
\label{5-6}\\
& &\delta{\ddot \eta}+\left(-8c_0 \pm 6\sqrt{-c_0c_1}\ \delta Q
\right)\delta\eta \approx \mp 6\sqrt[4]{-\frac{c_0c_1^2}{8}}\delta Q \ , 
\label{5-7}
\end{eqnarray}
where we used the static solutions in (\ref{5-2}). 
The solution of (\ref{5-6}) gives the small oscillation around 
the static configuration : 
\begin{equation}\label{5-8}
\delta Q=\sigma \cos (\sqrt{-2c_0}\ t+\theta_0) \ . 
\end{equation}
If we adopt the static solutions as $Q_0=+\sqrt{-c_0/c_1}$ and $\eta_0=+
\sqrt[4]{-1/8c_0}$, the equation for the quantum fluctuation 
$\eta$ are reduced to 
\begin{eqnarray}\label{5-9}
& &\delta{\ddot \eta}+\left(\! -8c_0+6\sigma\sqrt{-c_0c_1}
\cos\! \left(\sqrt{-2c_0}\ t+\theta_0\!\right)\!
\right)\delta\eta =-6 \sqrt[4]{-\frac{c_0c_1^2}{8}}
\sigma\cos \left(\sqrt{-2c_0}\ t+\theta_0\right) \ . \nonumber\\
& &
\end{eqnarray}
Adopting $\theta_0=\pi$ without the loss of generality, 
the above equation can be expressed as 
\begin{subequations}\label{5-10}
\begin{equation}
\delta{\ddot \eta}+\omega_0^2\left(1-h\cos \gamma t \right)\delta\eta 
=f\cos \gamma' t \ , 
\label{5-10a}
\end{equation}
where 
\begin{eqnarray}\label{5-10b}
& &\omega_0^2=-8c_0 \ , \quad
h=\frac{3}{4}\sigma\sqrt{-\frac{c_1}{c_0}} \ , \quad
f=6\sqrt[4]{-\frac{c_0c_1^2}{8}}\sigma \ , \quad
\gamma=\gamma'=\sqrt{-2c_0} \ .\qquad\ \ 
\end{eqnarray}
\end{subequations}
If the right-hand side of (\ref{5-10a}) or (\ref{5-9}) can be neglected, 
this equation is identical with the Mathieu equation. 
Thus, a solution which reveals the parametric resonance exists. 
On the other hand, if the second term of the bracket on the left-hand side 
in (\ref{5-10a}) or (\ref{5-9}) can be neglected, it is possible that 
the solution of this equation shows 
the forced oscillation.

\section{Discussion and concluding remarks}

The equation (\ref{5-9}), or equivalently (\ref{5-10a}), describes the 
time-evolution of quantum fluctuation around the collective motion. 
Here, $\delta\eta$ represents the time-dependent part of 
quantum fluctuation around the static configuration which is denoted 
as $\eta_0$. 
In this coupled Lipkin model, the amplification of the amplitude of 
quantum fluctuation, $\delta\eta$, may be induced 
by the collective oscillation, which 
is represented in terms of $\delta Q$, around 
the static configuration $Q_0$. 
Equation (\ref{5-10a}) reveals two possibility for the amplification of 
quantum fluctuation around the collective motion : One is by 
the resonance mechanism 
in the forced oscillation and the other is by 
the parametric resonance mechanism governed by the Mathieu equation. 
The amplification of the amplitude of quantum fluctuation inversely means 
the damping of collective mode.

If $h$ in Eq.(\ref{5-10a}) can be neglected, this equation of motion 
for quantum fluctuation is reduced to the 
well-known equation of forced oscillation in the classical 
mechanics.\cite{LL} 
However, $\omega_0=2\gamma'$ is satisfied as is seen in (\ref{5-10b}), 
it is not expected to realize the resonance 
phenomena proportional to $t$ 
originated by the forced oscillation, while the beat may occur.

On the other hand, if $f$ in (\ref{5-10a}) is neglected or the right-hand side 
in (\ref{5-10a}) has no effect, Eq.(\ref{5-10a}) can be reduced to the 
famous Mathieu equation. 
This can be realized even when the 
amplitude of collective oscillation, $\sigma$, is small. 
The Mathieu equation\cite{WW}
\begin{equation}\label{6-1}
\delta{\ddot \eta}+\omega_0^2\left(1-h\cos \gamma t \right)\delta\eta 
=0
\end{equation}
describes a parametric resonance phenomena. 
The parametric resonance occurs in the region around $\gamma=2\omega_0/n$ where $n$ is natural integer. When we denote $\gamma=2\omega_0/n+\varepsilon$ 
where $\varepsilon \ll 1$ and $h \ll 1$, 
the parametric resonance occurs in the region $-{\cal O}(h^n)<\varepsilon
<{\cal O}(h^n)$.\cite{LL} 
In this coupled Lipkin model in Eq.(\ref{5-10a}), 
$\omega_0=2\gamma$ 
is realized which corresponds to the case $n=4$ and $\varepsilon=0$ 
in the Mathieu equation (\ref{6-1}). 
Thus, the parametric resonance occurs inevitably, which may lead to the 
damping of the collective motion because of the growing of fluctuation energy 
induced by the amplification of the amplitude of quantum fluctuation around 
the collective variable. 
However, the damping time may be rather long because the parametric 
resonance works weakly because of $n=4$.

It should be noted here the possible scenario for the amplification 
of quantum fluctuation modes obtained from the 
results in this paper and the previous work in Ref.\citen{T02}. 
In the coupled Lipkin model, 
it has been indicated that the parametric resonance occurs 
for the quantum fluctuation mode around the collective motion. 
However, the instability of this fluctuation mode is weak 
because of $n=4$ in terms of the usual Mathieu equation in (\ref{6-1}). 
If the other fluctuation modes exist, the unstable modes with the 
lower $n$ such as $n=1$ may appear. 
This situation has actually been seen in the O(4) linear sigma model 
in the quantum field theory with sigma meson and pions.\cite{T02} 
In that case, the chiral condensate corresponds to the collective variable and 
the sigma meson modes correspond to the quantum fluctuation mode
around the collective motion. 
In the O(4) linear sigma model, amplification of quantum meson mode with 
$n=4$ in the Mathieu equation 
has been also realized in the lowest quantum 
sigma meson mode around the chiral condensate. 
This phenomenon is identical with that seen in the coupled Lipkin model 
investigated in this paper. In addition to the sigma meson modes, 
in the O(4) linear sigma model, the quantum pion modes exist. 
Then, it has been shown that 
the low momentum pion modes become unstable modes with $n=1$ 
in terms of the Mathieu equation. The amplification is strong compared 
with $n=4$ case in the sigma direction around the chiral condensate.   
Thus, if the other modes, such as the intrinsic modes, 
except for the collective mode and the quantum fluctuation mode around it 
exist and they are coupled with collective mode, the strong unstable modes 
may be realized and the dissipation of collective motion may occur. 
It is interesting to investigate the possibility of the dissipation 
of collective motion. 
In addition to the parametric resonance mechanism, the resonance by the 
forced oscillation may occur, as was seen in the O(4) linear sigma model. 
These are further problems to study.

In summary, the collective oscillation in the coupled Lipkin model 
inevitably leads to the amplification of the amplitude of 
quantum fluctuation mode around the collective mode. 
When the amplitude of the collective oscillation is small, 
the parametric resonance mechanism works to amplify the 
amplitude of quantum fluctuation mode, which may lead to damping of 
the collective oscillation.

\acknowledgement 
This work was completed when two of the authors (Y.T. and M.Y.) 
stayed at Coimbra in September 2004. They would like to 
express their sincere thanks to Professor Jo\~ao da Provid\^encia, 
co-author of this paper, for his warm hospitality. 
One of the authors (Y.T.) 
is partially supported by the Grants-in-Aid of the Scientific Research 
No.15740156 from the Ministry of Education, Culture, Sports, Science and 
Technology in Japan.

%\appendix
%\section{}


\begin{thebibliography}{99}
%%%%%%%%%%%%%%%%%%%%%%%%%%%%%%%%%%%%%%%%%%%%%%%%%%%%%%%%%%%%%
% Some macros are available for the bibliography:
%   o for general use
%      \JL : general journals          \andvol : Vol (Year) Page
%   o for individual journal 
%      \PR  : Phys. Rev.               \PRL : Phys. Rev. Lett.
%      \NP  : Nucl. Phys.              \PL  : Phys. Lett.
%      \JMP : J. Math. Phys.           \CMP : Commun. Math. Phys.
%      \PTP : Prog. Theor. Phys.       \JPSJ: J. Phys. Soc. Jpn.
%      \JP  : J. of Phys.              \NC  : Nouvo Cim.
%      \IJMP: Int. J. Mod. Phys.       \ANN : Ann. of Phys.
% Usage:
%   \PR{D45,1990,345}            ==> Phys.~Rev.\ {\bf D45} (1990), 345
%   \JL{Phys.~Lett.,A30,1981,56} ==> Phys.~Lett.\ {\bf A30} (1981), 56
%   \andvol{B123,1995,1020}      ==> {\bf B123} (1995), 1020
%%%%%%%%%%%%%%%%%%%%%%%%%%%%%%%%%%%%%%%%%%%%%%%%%%%%%%%%%%%%%
\bibitem{MMSK80}
T. Marumori, T. Maskawa, F. Sakata and A. Kuriyama, 
Prog. Theor. Phys. {\bf 64} (1980), 1294. 
\bibitem{KNMM04}
M. Kobayashi, T. Nakatsukasa, M. Matsuo and K. Matsuyanagi, 
%Prog. Theor. Phys. {\bf 112} (2004), 363; 
Prog. Theor. Phys. {\bf 113} (2005), 129. 
\bibitem{YK87sup}
M. Yamamura and A. Kuriyama, Prog. Theor. Phys. Suppl. No.93 (1987), 1, and 
references cited therein. 
\bibitem{KY84}
A. Kuriyama and M. Yamamura, Prog. Theor. Phys. 
{\bf 71} (1984), 122.
\bibitem{KYI84}
A. Kuriyama, M. Yamamura and S. Iida, Prog. Theor. Phys. 
{\bf 72} (1984), 1273; ibid.{\bf 73} (1985), 837. 
\bibitem{HI80}
A. Hayashi and S. Iwasaki, Prog. Theor. Phys. {\bf 63} (1980), 1063. 
\bibitem{MM}
S. Mr\'owczy\'nski and B. M\"uller, 
        Phys.~Lett.~{\bf B363} (1995), 1.  
\bibitem{HM}
H. Hiro-Oka and H. Minakata, 
        Phys.~Rev.~{\bf C61} (2000), 044903; ibid.{\bf C64} (2001), 044902.
\bibitem{T02}
Y. Tsue, Prog. Theor. Phys. {\bf 107} (2002), 1285. 
\bibitem{BS03}
J. Berges and J. Serreau, Phys. Rev. Lett. {\bf 91} (2003), 111601.
\bibitem{KY83}
A. Kuriyama and M. Yamamura, Prog. Theor. Phys. 
{\bf 69} (1983), 681.
\bibitem{TFKY91}
Y. Tsue, Y. Fujiwara, A. Kuriyama and M. Yamamura, Prog. Theor. Phys. 
{\bf 85} (1991), 693.
\bibitem{TF91}
Y. Tsue and Y. Fujiwara, Prog. Theor. Phys. 
{\bf 86} (1991), 443.
\bibitem{KPTYsuppl}
A. Kuriyama, J. da Provid\^encia, Y. Tsue and M. Yamamura, 
Prog. Theor. Phys. Suppl. No. 141 (2001), 113. 
\bibitem{LL}
L. D. Landau and E. M. Lifshitz,         
        {\it Mechanics}, 
        (Butterworth Heinemann, 1976), \S 27.
\bibitem{WW}
E. T. Whittaker and G. N. Watson, 
        {\it A course of Modern Analysis}, 
        (Cambridge at the University Press, 1935), p.404.
\end{thebibliography}
\end{document}